\documentclass[sigconf]{acmart}
\acmConference[MSR 2026]{23rd International Conference on Mining Software Repositories}{April 2026}{Rio de Janeiro, Brazil}

\usepackage[english]{babel}
\usepackage[]{color}
\usepackage{listings}
\usepackage{xcolor}
\usepackage{colortbl}
\usepackage{comment}
\usepackage{graphicx}
\usepackage{array, colortbl}
\usepackage{listings}
\usepackage{multirow}
\usepackage{rotating}
\usepackage{float}
\usepackage[]{url}
\usepackage{balance}
\usepackage{fancybox}
\usepackage{scalefnt}
\usepackage{subcaption}
\usepackage{cleveref}
\usepackage{xspace}
\usepackage{caption}
\usepackage{changepage}
\usepackage{booktabs}
\usepackage{diagbox}
\usepackage{natbib}
\usepackage{tabularx}
\usepackage{enumitem}
\usepackage{url}
\usepackage{fnpct}
\usepackage{caption}
\setlist[enumerate]{leftmargin=*}

\definecolor{amber}{rgb}{1.0, 0.49, 0.0}
\definecolor{lavenderindigo}{rgb}{0.58, 0.34, 0.92}
\definecolor{islamicgreen}{rgb}{0.0, 0.56, 0.0}

\definecolor{LightCyan}{rgb}{0.88,1,1}
\definecolor{beaublue}{rgb}{0.74, 0.83, 0.9}
\definecolor{bubbles}{rgb}{0.91, 1.0, 1.0}

\definecolor{ScarletRed}{rgb}{0.80,0.00,0.00}

\makeatletter
\newcommand{\labelname}[1]{%
  \def\@currentlabelname{#1}}%
\makeatother

\usepackage{xcolor}

\colorlet{punct}{red!60!black}
\definecolor{background}{HTML}{EEEEEE}
\definecolor{delim}{RGB}{20,105,176}
\colorlet{numb}{magenta!60!black}

\AtBeginDocument{%
  \providecommand\BibTeX{{%
    \normalfont B\kern-0.5em{\scshape i\kern-0.25em b}\kern-0.8em\TeX}}}

\setcopyright{acmlicensed}
\copyrightyear{2026}
\acmYear{2026}
\acmDOI{XXXXXXX.XXXXXXX}
\acmConference[MSR '26]{Make sure to enter the correct
  conference title from your rights confirmation email}{April 13--14,
  2026}{Rio de Janeiro, Brazil}

\begin{document}

\title{Tokenomics: Quantifying Where Tokens Are Used in Agentic Software Engineering}

\author{Mohamad Salim}
\affiliation{%
\institution{Data-driven Analysis of Software (DAS) Lab \\ Concordia University}
\city{Montreal}
\country{Canada}}
\email{mo_alim@live.concordia.ca}

\author{Jasmine Latendresse}
\affiliation{%
\institution{Data-driven Analysis of Software (DAS) Lab \\ Concordia University}
\city{Montreal}
\country{Canada}}
\email{jasmine.latendresse@mail.concordia.ca}

\author{SayedHassan Khatoonabadi}
\affiliation{%
\institution{Data-driven Analysis of Software (DAS) Lab \\ Concordia University}
\city{Montreal}
\country{Canada}}
\email{sayedhassan.khatoonabadi@concordia.ca}

\author{Emad Shihab}
\affiliation{%
\institution{Data-driven Analysis of Software (DAS) Lab \\ Concordia University}
\city{Montreal}
\country{Canada}}
\email{emad.shihab@concordia.ca}

\begin{abstract}
LLM-based Multi-Agent (LLM-MA) systems are increasingly applied to automate complex software engineering tasks such as requirements engineering, code generation, and testing. However, their operational efficiency and resource consumption remain poorly understood, hindering practical adoption due to unpredictable costs and environmental impact. To address this, we conduct an analysis of token consumption patterns in an LLM-MA system within the Software Development Life Cycle (SDLC), aiming to understand where tokens are consumed across distinct software engineering activities. We analyze execution traces from 30 software development tasks performed by the ChatDev framework using a GPT-5 reasoning model mapping its internal phases to distinct development stages (Design, Coding, Code Completion, Code Review, Testing, and Documentation) to create a standardized evaluation framework. We then quantify and compare token distribution (input, output, reasoning) across these stages.

Our preliminary findings show that the iterative Code Review stage accounts for the majority of token consumption for an average of 59.4\% of tokens. Furthermore, we observe that input tokens consistently constitute the largest share of consumption for an average of 53.9\%, providing empirical evidence for potentially significant inefficiencies in agentic collaboration. Our results suggest that the primary cost of agentic software engineering lies not in initial code generation but in automated refinement and verification. Our novel methodology can help practitioners predict expenses and optimize workflows, and it directs future research toward developing more token-efficient agent collaboration protocols.
\end{abstract}


\maketitle

\section{Introduction}
\label{sec:introduction}
Large-scale software engineering increasingly explores LLM-Based Multi-Agent (LLM-MA) systems to automate complex tasks across the Software Development Life Cycle (SDLC) \cite{he2025llm, lu2025exploring}. These LLM-MA frameworks simulate human teams (e.g., product managers, architects, developers, testers) using specialized Large Language Model (LLM) agents that collaborate to design, code, and verify software. In principle, LLM-MA systems can improve autonomy and robustness by dividing work across agents \cite{he2025llm}. Prior work highlights that LLM-MA systems encourage divergent thinking \cite{liang2024encouraging}, enhance reasoning and factuality \cite{du2024improving}, and scale to problems beyond single-agent capacity \cite{he2025llm}. For Software Engineering (SE), this suggests LLM-MA systems could automate end-to-end workflows, from requirements to testing, in a unified manner \cite{hong2024metagpt, qian2024chatdev}.

Recent studies have begun to analyze the behavior and efficiency of these systems. The AGENTTAXO framework \cite{wang2025agenttaxo} provided a taxonomy for dissecting token distribution in general LLM-MA systems, introducing the concept of a "communication tax" to describe the overhead from inter-agent interactions. In addition, the MAST taxonomy of failures revealed that many issues in LLM-MA systems stem from systemic design and coordination challenges, such as step repetition or incomplete verification, rather than individual LLM limitations \cite{pan2025why}. While this prior work establishes essential taxonomies for understanding token distribution and systemic failure modes, it analyzes agent behavior in a general context. A significant knowledge gap exists regarding the resource efficiency of these systems when applied specifically to the unique, multi-stage SE workflows \cite{qiu2025co}. The ultimate question for practical adoption—"Where do the tokens go?"—remains unanswered in the SE domain. 

Thus, in this paper, we introduce the term "tokenomics" as the study of operational efficiency and resource consumption in LLM-MA systems. To our knowledge, this is the first study to conduct an empirical analysis of tokenomics in an SE context that examines execution traces of an SE-focused LLM-MA system through the lens of the SDLC. To guide our study, we focus on the following fundamental research question: \textbf{What are the token consumption patterns of LLM-MA systems for software development tasks?} 

To answer this, we analyze the distribution of token consumption across distinct development stages, which we derive by mapping the internal phases of the multi-agent framework \textit{ChatDev} \cite{qian2024chatdev}. ChatDev simulates a virtual software company where multiple agent roles (e.g., programmer, tester) collaborate through multi-turn dialogues to complete the SDLC. Answering this question is the first step towards building economically and environmentally sustainable agentic SE systems.

This paper contributes an empirical analysis, a curated dataset of 30 execution traces, and a complete replication package\footnote{\url{https://zenodo.org/records/17430187}}.

\begin{figure*}
    \centering
    \includegraphics[width=\textwidth]{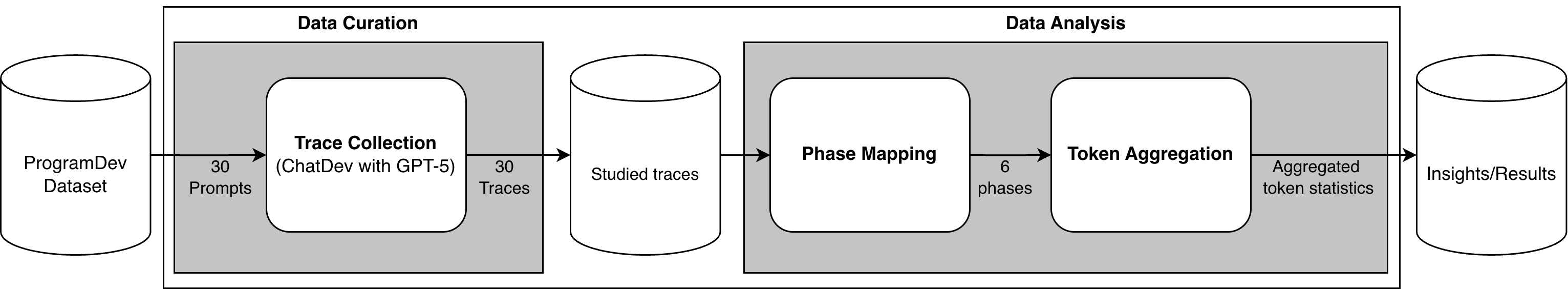}
    \caption{Overview of our analysis pipeline.}
    \label{fig:pipeline_overview}
\end{figure*}

\section{Study Design}
\label{sec:design}
The goal of our study is to empirically investigate the distribution of token consumption within an LLM-MA system as it performs end-to-end software development tasks. To achieve this, we selected \textit{ChatDev} \cite{qian2024chatdev} as our initial system for analysis. We make this choice because its "chat chain" architecture represents a clear, sequential waterfall model (design $\rightarrow$ coding $\rightarrow$ testing), making its phases distinct and well-suited for mapping to software development stages. In addition, this framework is one of the most popular and highly cited open source frameworks \cite{wang2024survey, guo2024large, qiu2025co}.

\subsection{Dataset Curation}
\label{sec:dataset}
We executed ChatDev \cite{qian2024chatdev} on 30 distinct software development tasks, with the prompts sourced from the ProgramDev Dataset \cite{pan2025why}, which was used in the foundational MAST study \cite{pan2025why}. The selected prompts range from simple algorithms (e.g., Fibonacci number generation) to more complex applications (e.g., a chess game), ensuring task diversity. Recent work suggests that the number of reasoning tokens allocated by a model can serve as a proxy for task complexity \cite{shojaee2025illusion}. Our dataset exhibits a wide range in reasoning tokens consumed across the 30 tasks (from 17,280 to 40,000 tokens), which suggests a sufficient diversity in task complexity for this study. 

\subsection{Model Selection}
\label{sec:model}
The GPT-5 reasoning model was selected as the backbone for all agents. This decision was based on the popularity and recency of the model, its suitability for agentic use cases, and its strong reasoning capabilities, which are in line with the expectations of autonomous agents \cite{he2025llm}.
As detailed in Table~\ref{tab:gpt5_config}, the model version used is \textit{gpt-5-2025-08-07}. The temperature parameter is not supported for this model, so the default value \textit{1.0} was used. 
% Its context window is \textit{400,000 tokens} and its max output tokens is \textit{128,000 tokens}. The model's knowledge cutoff date is \textit{Sep 30, 2024.} 

\begin{table}
    \centering\small
    \caption{Details of the GPT-5 model used in the experiments.}
    \label{tab:gpt5_config}
    \begin{tabular}{ll}
        \toprule
        \textbf{Parameter} & \textbf{Value} \\
        \midrule
        Model Version & gpt-5-2025-08-07 \\
        Temperature & 1.0 (default value; immutable) \\
        Context Window & 400,000 tokens \\
        Max Output Tokens & 128,000 tokens \\
        Knowledge Cutoff & Sep 30, 2024 \\
        \bottomrule
    \end{tabular}
\end{table}

\subsection{Analysis Pipeline}
\label{sec:analysis}
To analyze the collected data, we designed and implemented a multi-step pipeline, illustrated in Figure~\ref{fig:pipeline_overview}.

\noindent \textbf{Trace Collection.} We instrumented ChatDev to log the complete execution trace for each of the 30 tasks, capturing every LLM call, including the prompt, response, and associated token counts (input, output, reasoning).

\noindent \textbf{Phase Mapping.} A core methodological contribution of our work is the mapping of ChatDev's internal, framework-specific phases to universally understood development stages. This abstraction allows for generalizable analysis and can be extended to other SE LLM-MA frameworks. The mapping used is detailed in Table~\ref{tab:chatdev_mapping}.

\noindent \textbf{Token Aggregation.} Using this mapping, we wrote Python scripts to parse the traces collected and aggregate token counts for each of the development stages across all 30 runs, calculating totals and breaking them down by input, output, and reasoning tokens.

\begin{table*}
    \centering\footnotesize
    \caption{Mapping of ChatDev's internal phases to distinct software development stages. The mapping is based on the descriptions of each phase provided in the framework's documentation.}
    \label{tab:chatdev_mapping}
    \begin{tabularx}{\textwidth}{l>{\raggedright}X>{\raggedright\arraybackslash}X}
        \toprule
        \textbf{Development Stage} & \textbf{ChatDev Phases} & \textbf{Description} \\
        \midrule
        Design & DemandAnalysis, LanguageChoose & These initial phases focus on understanding requirements and making high-level technical decisions. \\
        \addlinespace
        Coding & Coding & This phase is directly involved in writing the initial source code. \\
        \addlinespace
        Code Completion & CodeComplete & This phase completes any placeholder or incomplete code files left from the Coding phase. \\
        \addlinespace
        Code Review & CodeReview & This phase involves an iterative dialogue between a programmer and code reviewer agent to review and modify/refine code. \\
        \addlinespace
        Testing & Test & This phase explicitly focuses on dynamic system testing to locate and fix executability bugs. \\
        \addlinespace
        Documentation & EnvironmentDoc, Reflection, Manual & These final phases generate user manuals and document required environment dependencies. \\
        \bottomrule
    \end{tabularx}
\end{table*}

\section{Study Results}
\label{sec:results}
In this section, we present the results of our research question. We present its motivation, the approach to answer the question, and the results.

\subsection{RQ: What are the token consumption patterns of LLM-MA systems for software development tasks?}
\label{sec:rq}

\noindent \textbf{Motivation.} Understanding the token consumption patterns, or "tokenomics," of agentic SE systems is critical for their practical and sustainable adoption. High token usage translates directly to increased financial costs, energy consumption, and environmental impact. By identifying where tokens are consumed within the SDLC, we can create a "cost map" that enables practitioners to predict expenses and optimize workflows. While prior work has analyzed general MAS behavior, there is a clear gap in understanding these efficiency patterns specifically within the context of software development \cite{qiu2025co}, which this RQ aims to address.

\noindent \textbf{Approach.} To answer this question, we analyze the aggregated token data from the study pipeline described in Section~\ref{sec:design}. We focus on two primary dimensions:

\begin{enumerate}
    \item The distribution of total tokens across the mapped development stages (Design, Coding, etc.)
    \item The ratio of input, output, and reasoning tokens within each stage.
\end{enumerate}

\begin{figure}[]
    \centering
    \includegraphics[width=\columnwidth]{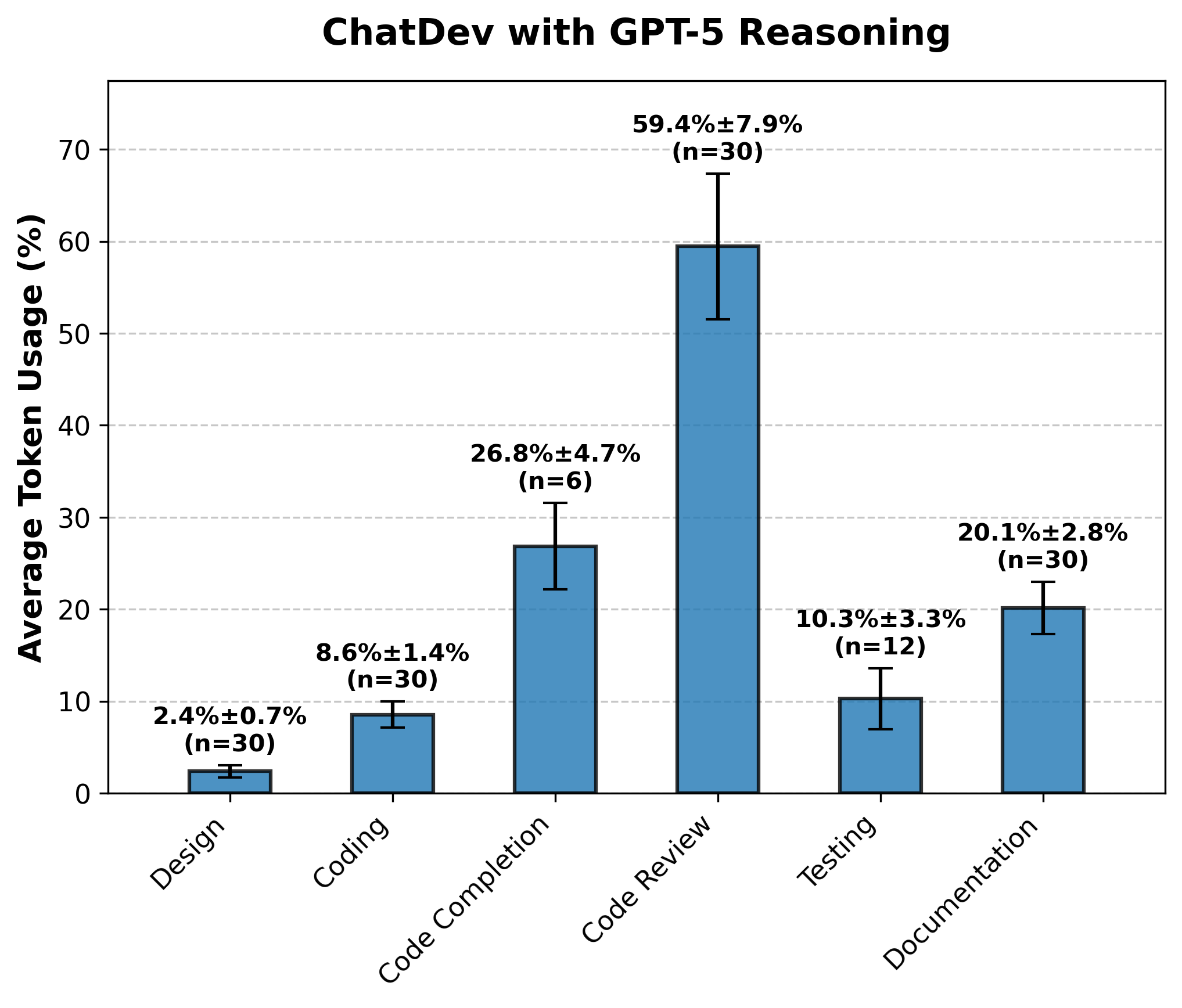}
    \caption{Average token usage by phase across all 30 tasks for ChatDev with GPT-5 Reasoning. Error bars represent $\pm1$ standard deviation. Note that not all phases occurred in every task (e.g., Code Completion n=6, Testing n=12).}
    \label{fig:Bar_Chart_ChatDev_GPT-5}
\end{figure}

\noindent \textbf{Finding 1: The Code Review Stage Dominates Token Consumption.}
Our analysis reveals a highly uneven distribution of token usage across the development process. As shown in Figure~\ref{fig:Bar_Chart_ChatDev_GPT-5}, a clear hierarchy of token consumption emerges. In the figure, the "n" value denotes the number of tasks (out of 30) where a specific phase was executed. This value is not always "30", as the agents within the multi-agent system autonomously decide which phases to execute, and not all phases are needed for every task. The error bars represent $\pm$1 standard deviation, indicating the variability in token consumption for each phase. \textbf{The Code Review phase is the largest consumer, responsible for an average of 59.4\% of tokens across all 30 tasks.} The Code Completion phase, which occurred in 6 of the 30 tasks, was also costly, averaging $26.8\%$ of tokens in those runs. These two refinement-focused stages are followed by Documentation (avg. 20.1\%) and Testing (avg. 10.3\%), the latter of which occurred in 12 of the 30 tasks. In contrast, initial Coding (avg. 8.6\%) and Design (avg. 2.4\%) are remarkably inexpensive. This suggests that the primary cost of agentic software engineering lies not in initial code generation but in the iterative, conversational process of refinement and verification.

\noindent \textbf{Finding 2: Token Consumption is Dominated by Input Tokens.} Across all phases except the Coding phase, we observe a consistent pattern where \textbf{input tokens far exceed output and reasoning tokens.} On average, the overall token usage for each task analyzed is composed of $53.9\%$ input tokens, $24.4\%$ output tokens, and $21.6\%$ reasoning tokens. This \textbf{approximate 2:1 ratio of input to output tokens provides strong empirical evidence for the "communication tax"} identified in prior work \cite{wang2025agenttaxo}, where agents repeatedly pass large contexts during their collaborative dialogue. This highlights a significant inefficiency in current agent collaboration protocols, where the majority of the tokens are spent on communicating context rather than generating novel output. This also suggests that the communication tax may be an inherent characteristic of conversational multi-agent architectures, a phenomenon which future work should investigate further.

\noindent \textbf{Finding 3: Software Development Stages Exhibit Distinct Tokenomic Profiles.} A deeper look at the token ratios per phase, detailed in Table~\ref{tab:token_ratio_breakdown}, reveals unique patterns for different software engineering activities. The Coding phase is a notable outlier, being output-heavy (58\% output vs. 6.9\% input). This is intuitive, as it involves generating verbose source code from a more concise design specification. In contrast, verification phases like Code Review, and documentation phases are input-heavy (51.4\% and 80.2\% input, respectively). These phases consume large amounts of existing code as context to produce small, analytical outputs. These distinct profiles provide a "cost map" for different engineering activities, enabling practitioners to better predict expenses and identify opportunities for process optimization.

\begin{table}
    \centering\footnotesize
    \caption{Phase-by-phase token ratio breakdown for ChatDev with GPT-5 Reasoning across 30 tasks.}
    \label{tab:token_ratio_breakdown}
    \begin{tabular}{l|ccc}
        \toprule
        \textbf{Development Stage} & \textbf{Avg Input} \% & \textbf{Avg Output} \% & \textbf{Avg Reasoning} \% \\
        \midrule
        Design & \textbf{60.4} & 3.6 & 36.0 \\
        Coding & 6.9 & \textbf{58.0} & 35.1 \\
        Code Completion & \textbf{47.7} & 41.7 & 10.5 \\
        Code Review & \textbf{51.4} & 24.7 & 23.9 \\
        Testing & \textbf{60.8} & 20.7 & 18.4 \\
        Documentation & \textbf{80.2} & 8.3 & 11.5 \\
        \midrule
        Overall (per task) & \textbf{53.9} & 24.4 & 21.6 \\
        \bottomrule
    \end{tabular}
\end{table}

\vspace{0.2cm}
\noindent\fcolorbox{black}{gray!20!white}{\parbox{\columnwidth}{
    \textbf{Answer to RQ:} With the \textit{ChatDev} LLM-MA system, token consumption is heavily concentrated in the Code Review stage of the SDLC. Moreover, token consumption is dominated by input tokens, reflecting a potentially significant communication tax, with different development stages exhibiting unique tokenomic profiles corresponding to the nature of the software engineering task (e.g., planning, reasoning, or verification).
}}

\section{Discussion}
\label{sec:discussion}

Our preliminary results offer an initial "cost map" of agentic software development, with several implications for practitioners and researchers.

The immense token cost of the Code Review phase can be interpreted as the "Cost of Conversation." This is a direct consequence of the inherent conversational architecture of LLM-MA systems, where agents iteratively pass the full code context back and forth to refine it. \textbf{This suggests that current agentic collaboration protocols for verification are highly inefficient, consuming vast resources to perform tasks that might involve minor corrections.} This aligns with findings from the MAST taxonomy \cite{pan2025why}, where failures related to verification and step repetition are common, suggesting that high token usage may be a symptom of the agentic system's attempt to overcome these inherent coordination challenges through brute-force dialogue.   

For practitioners, our findings provide a basis for cost prediction and process optimization. The distinct tokenomic profiles imply that \textbf{the cost of an agent-driven project can be estimated based on the type of work required.} For example, greenfield projects with heavy initial coding will have a different cost structure than projects focused on refactoring and debugging existing code, which will be dominated by the expensive, input-heavy code review cycle. This insight can inform design decisions, such as integrating a "human-in-the-loop" checkpoint before the Code Review phase to prevent costly iterative loops \cite{navneet2025rethinking}, thereby maximizing both economic and computational efficiency.

For the research community, our results present a clear challenge and a potential solution. The challenge is \textbf{to design more token-efficient collaboration protocols for verification and refinement, moving beyond naive full-context passing.} In addition, there is a clear \textbf{need for a standardized, comprehensive evaluation framework \cite{li2025beyond}.} This framework can serve as a common ground to benchmark and compare the efficiency of different LLM-MA architectures (e.g., ChatDev's hierarchical, conversational workflow vs. MetaGPT's SOP-based assembly line) in future work, providing a "Rosetta Stone" to translate framework-specific operations into universal software engineering activities.

\section{Threats to Validity}
\label{sec:threats_to_validity}
There are a few important limitations to our work that need to be considered when interpreting our findings. Firstly, our analysis is based on a single LLM-MA system (ChatDev) and a single LLM (GPT-5 Reasoning Model). The observed token consumption patterns may differ in other LLM-MA architectures or with other LLMs that have different token efficiencies \cite{wang2025agenttaxo}. Secondly, the 30 software development tasks, while diverse, may not represent all possible software development scenarios and complexities. The size of the curated dataset is a direct consequence of the current lack of public, large-scale benchmarks for SE-specific agent traces \cite{yan2025beyond}, which makes data curation a time-consuming and costly process. Thirdly, some development stages were executed in only a small subset of the 30 tasks. For instance, the Code Completion ($n=6$) and Testing ($n=12$) phases were triggered infrequently by the agentic system. The conclusions drawn about the tokenomic profiles of these specific stages are based on a small sample, which may not be representative and may limit the generalizability of those particular findings. Finally, our proposed mapping of ChatDev's internal phases to software development stages is an abstraction. While we believe it is a logical and useful one for creating a standardized evaluation framework, it represents one of several possible mappings of the agent's activities.

\section{Conclusion and Future Work}
\label{sec:conclusion} 
This work-in-progress paper sets out to answer \textit{"where do the tokens go?"} in agentic software engineering. Our preliminary empirical study using the \textit{ChatDev} framework reveals that the answer is not straightforward. The costs are not evenly distributed but are overwhelmingly concentrated in the iterative, conversational phase of code review. We also found that input tokens, comprising the "communication tax," form the bulk of the token usage, highlighting a key area for future optimization.

This study lays the groundwork for a comprehensive research agenda. Future work should focus on: 
\begin{enumerate}
    \item Expanding our dataset with more tasks to ensure better generalizability.
    \item Extending the analysis to other LLMs to understand model-specific effects.
    \item Extending the analysis to other LLM-MA systems to perform a comparative study of how architectural differences impact tokenomics.
    \item Investigating the relationship between token consumption patterns and failure modes.
    \item Further developing and validating our development stage mapping as a robust, universal framework for benchmarking SE agent efficiency.
\end{enumerate}

%\nocite{*}
\bibliographystyle{ACM-Reference-Format}
\bibliography{main.bib}

\end{document}